\PassOptionsToPackage{unicode}{hyperref}
\PassOptionsToPackage{hyphens}{url}
\documentclass[
]{article}
\usepackage{lmodern}
\usepackage{amssymb,amsmath}
\usepackage{ifxetex,ifluatex}
\ifnum 0\ifxetex 1\fi\ifluatex 1\fi=0 % if pdftex
  \usepackage[T1]{fontenc}
  \usepackage[utf8]{inputenc}
  \usepackage{textcomp} % provide euro and other symbols
\else % if luatex or xetex
  \usepackage{unicode-math}
  \defaultfontfeatures{Scale=MatchLowercase}
  \defaultfontfeatures[\rmfamily]{Ligatures=TeX,Scale=1}
\fi
\IfFileExists{upquote.sty}{\usepackage{upquote}}{}
\IfFileExists{microtype.sty}{% use microtype if available
  \usepackage[]{microtype}
  \UseMicrotypeSet[protrusion]{basicmath} % disable protrusion for tt fonts
}{}
\makeatletter
\@ifundefined{KOMAClassName}{% if non-KOMA class
  \IfFileExists{parskip.sty}{%
    \usepackage{parskip}
  }{% else
    \setlength{\parindent}{0pt}
    \setlength{\parskip}{6pt plus 2pt minus 1pt}}
}{% if KOMA class
  \KOMAoptions{parskip=half}}
\makeatother
\usepackage{xcolor}
\IfFileExists{xurl.sty}{\usepackage{xurl}}{} % add URL line breaks if available
\IfFileExists{bookmark.sty}{\usepackage{bookmark}}{\usepackage{hyperref}}
\hypersetup{
  pdftitle={Spam four ways: Making sense of text data},
  pdfauthor={Nicholas J. Horton, Jie Chao, William Finzer, and Phebe Palmer},
  hidelinks,
  pdfcreator={LaTeX via pandoc}}
\usepackage[margin=1in]{geometry}
\usepackage{longtable,booktabs}
\usepackage{etoolbox}
\makeatletter
\patchcmd\longtable{\par}{\if@noskipsec\mbox{}\fi\par}{}{}
\makeatother
\IfFileExists{footnotehyper.sty}{\usepackage{footnotehyper}}{\usepackage{footnote}}
\makesavenoteenv{longtable}
\usepackage{graphicx}
\makeatletter
\def\maxwidth{\ifdim\Gin@nat@width>\linewidth\linewidth\else\Gin@nat@width\fi}
\def\maxheight{\ifdim\Gin@nat@height>\textheight\textheight\else\Gin@nat@height\fi}
\makeatother
\setkeys{Gin}{width=\maxwidth,height=\maxheight,keepaspectratio}
\makeatletter
\def\fps@figure{htbp}
\makeatother
\providecommand{\tightlist}{%
  \setlength{\itemsep}{0pt}\setlength{\parskip}{0pt}}
\title{Spam four ways: Making sense of text data}
\author{Nicholas J. Horton, Jie Chao, William Finzer, and Phebe Palmer}
\date{February 11, 2022}

\begin{document}
\maketitle

\hypertarget{abstract}{%
\subsection*{Abstract}\label{abstract}}
\addcontentsline{toc}{subsection}{Abstract}

The world is full of text data, yet text analytics has not traditionally played a large part in statistics education.
We consider four different ways to provide students with opportunities to explore whether email messages are unwanted correspondence (spam).
Text from subject lines are used to identify features that can be used in classification.
The approaches include use of a Model Eliciting Activity, exploration with CODAP, modeling with a specially designed Shiny app, and coding more sophisticated analyses using R.
The approaches vary in their use of technology and code but all share the common goal of using data to make better decisions and assessment of the accuracy of those decisions.

\hypertarget{introduction}{%
\subsection*{Introduction}\label{introduction}}
\addcontentsline{toc}{subsection}{Introduction}

The world is full of data, and much of it is unstructured. An example is text data, which forms a critical part of our lives through books, magazines, and the internet. Surprisingly, despite the key role of language arts in all aspects of education, text analysis has traditionally not played a major part in statistics education.

While there are many interesting literary analyses that one might consider, we explore a more mundane but familiar example by exploring when a text string taken from an email subject line is spam (an unwanted or inappropriate email message).

In this column we briefly describe four different ways to provide students with experiences classifying text as data by exploring spam. The data for the examples along with helpful links can be found in the GitHub repository accompanying this column at \href{http://bit.ly/taking-a-chance}{\emph{bit.ly/taking-a-chance}}.

\hypertarget{what-is-spam}{%
\subsection*{What is spam?}\label{what-is-spam}}
\addcontentsline{toc}{subsection}{What is spam?}

Email users are unfortunately all too familiar with spam.
The website \href{https://www.spamlaws.com}{\emph{www.spamlaws.com}}, based on materials created by David E. Sorkin from John Marshall Law School, provides some additional background on these indiscriminate and unwanted communications.
This site defines spam as ``submitting the same message to a large group of individuals in an effort to force the message onto people who would otherwise choose not to receive this message.''
The problem of spam is notable: they estimate that there are more than 14 billion spam messages created daily: a substantial fraction of all emails.

\hypertarget{spam-four-ways-approaches-to-teaching-students-to-classify-spam}{%
\subsection*{SPAM Four Ways: Approaches to teaching students to classify SPAM}\label{spam-four-ways-approaches-to-teaching-students-to-classify-spam}}
\addcontentsline{toc}{subsection}{SPAM Four Ways: Approaches to teaching students to classify SPAM}

To keep things tractable, we will focus on ways that one might use information from subject lines to determine whether emails are likely to be ``spam'' (unwanted messages) or ``non-spam''.
This is a form of \emph{supervised learning} or \emph{predictive analytics}, where the email subject lines are the data and the labels are the indicators of spam (or not) and our goal is predicting the outcome.
We realize that to some extent whether a message is considered spam or not is subjective: the question of what label is given to subject lines can engender a productive conversation.
This form of \emph{text analytics} is also a rudimentary type of \emph{natural language processing}, albeit with the simple goal of classification rather than the more challenging one of comprehension.

Five examples of the type of email subject lines that we would call spam include:

\begin{itemize}
\tightlist
\item
  Dear trusted one
\item
  From Mrs Kadirat Usman.thanks and remain bless. Urgent please
\item
  FBI \& IRS seized goods at 99\% off! Police Auctions!
\item
  Re: PROTECT YOUR COMPUTER AGAINST HARMFUL VIRUSES
\item
  Market Internet Access - No Investment Needed
\end{itemize}

Five examples of legitimate (non-spam) email subjects include:

\begin{itemize}
\tightlist
\item
  Receipt for your Payment to Edible Twin Cities
\item
  Your Zappos.com order \#: 65801179
\item
  STEM Education: faculty opening
\item
  Learning Outcomes working retreat
\item
  Re: Classifier software design
\end{itemize}

Before we proceed further, it's important to keep in mind (particularly for instructors) that as in any real-world investigation involving data from the internet, unsavory, obscene, or offensive email subject lines can arise.

\hypertarget{model-eliciting-activities-and-spam-classification}{%
\subsubsection*{Model Eliciting Activities and SPAM classification}\label{model-eliciting-activities-and-spam-classification}}
\addcontentsline{toc}{subsubsection}{Model Eliciting Activities and SPAM classification}

The CATALST (Change Agents for Teaching and Learning Statistics) Group defines Model Eliciting Activities (MEA) as those that are designed to let students explore open-ended problems that represent real-world problems.
We consider an MEA developed by researchers at the University of Minnesota where students are encouraged to invent approaches to classify spam messages.
This MEA has often been incorporated into an undergraduate introductory statistics course by the first author.

The activity is designed to fit within a 70-minute class period (though it could be split into two class sessions).
Students begin by reading a brief overview of the problem of spam and then individually complete some readiness questions (e.g., How can you determine the accuracy of a spam filter?).
The overall goal of the task is described as developing a spam filter for their work supervisor.

The students are divided into groups of between 2 and 5 students.
Each group is given two pieces of paper: one includes a list of 50 email subject lines labeled as ``spam'' and the other a list of 50 non-spam subject lines.
(The sample subject lines given earlier were taken from these lists.)

After reviewing the two sets of 50 subject lines, the students are asked to develop a rule or set of rules that will classify emails using only the subject line and to determine how they would judge the accuracy of their rule.
These rules might involve the development of word lists (e.g., ``Dear'', ``Blessed'', ``Urgent'') indicative of spam or others (e.g., ``Re'' or ``STEM'') that might be more common amongst non-spam.

Once the group comes to a consensus regarding their rules and apply it to the data they were provided, they are asked to write them down then share the rule with the class.
They were also asked to report on \emph{sensitivity} (proportion of spam that was declared to be spam) and \emph{specificity} (proportion of non-spam declared to be non-spam) of the rule on their training data in a table with rows for each group.

Rules vary considerably, with some relatively simple (e.g., if any of these words appear, call it spam) and others containing more complicated Boolean logic with multiple steps (e.g., if these words appear, declare it non-spam, else if more than two punctuation marks or majority capitalized then declare it spam).
It's been valuable to have each member of the team use the rule independently to calculate sensitivity and specificity to confirm that there is a shared understanding of how it is implemented.

The next step is for the teams to apply their rule to a set of new subject lines: 50 spam and 50 non-spam that are then provided.
They then report their results: it is not uncommon that their rules fare slightly worse on this testing data than on the original training data.

The final step in the MEA is for the group to write a short (1-2 page report) for their work supervisor on their results that includes a description of the method, their measures of accuracy, an assessment of how it worked (or not), and a way that they might adapt the spam filter based on their results.
Students draft the report outside of class.
This reflective assignment is intended to help make the concepts and process more concrete and to help students improve their ability to practice the language of statistics.

On its own, this MEA has felt like a valuable way to introduce aspects of predictive analytics using logistic regression or decision trees.
The activity helps students to engage in \emph{feature engineering} (also known as \emph{feature encoding}), an important step in machine learning.
It reinforces the idea that statistics can be used to make decisions, albeit with some uncertainty, and that various errors have different implications (e.g., seeing spam messages that were not flagged vs.~having a non-spam message marked as spam).
Choosing an example that students have familiarity with may help them feel more confident in their assessments, and therefore feel more comfortable and motivated to engage with the activity.

\hypertarget{classifying-spam-using-codap}{%
\subsubsection*{Classifying SPAM using CODAP}\label{classifying-spam-using-codap}}
\addcontentsline{toc}{subsubsection}{Classifying SPAM using CODAP}

One major advantage of the spam MEA is that no technology is needed, only paper (though typically teams will create a shared Google Drive file to draft and edit their report).
There are many ways to incorporate technology that allow students to further explore classification of subject lines.

We will begin by demonstrating two approaches that are available within CODAP (Common Online Data Analysis Platform) developed and maintained by the Concord Consortium.
CODAP is a free web-based environment for data analysis.
This environment is attractive for this application because it is free, has a simple and clean user interface, and runs in a browser.
CODAP documents can be distributed to students (or CHANCE readers!) by clicking ``Share'', which generates a URL that can allow another session to be started from a specific working state.
In recent years, a powerful plugin for feature creation and text classification has been developed through the ``Narrative Modeling with StoryQ'' project, an NSF funded project (see Further Resources).

\begin{figure}

{\centering \includegraphics[width=0.8\linewidth]{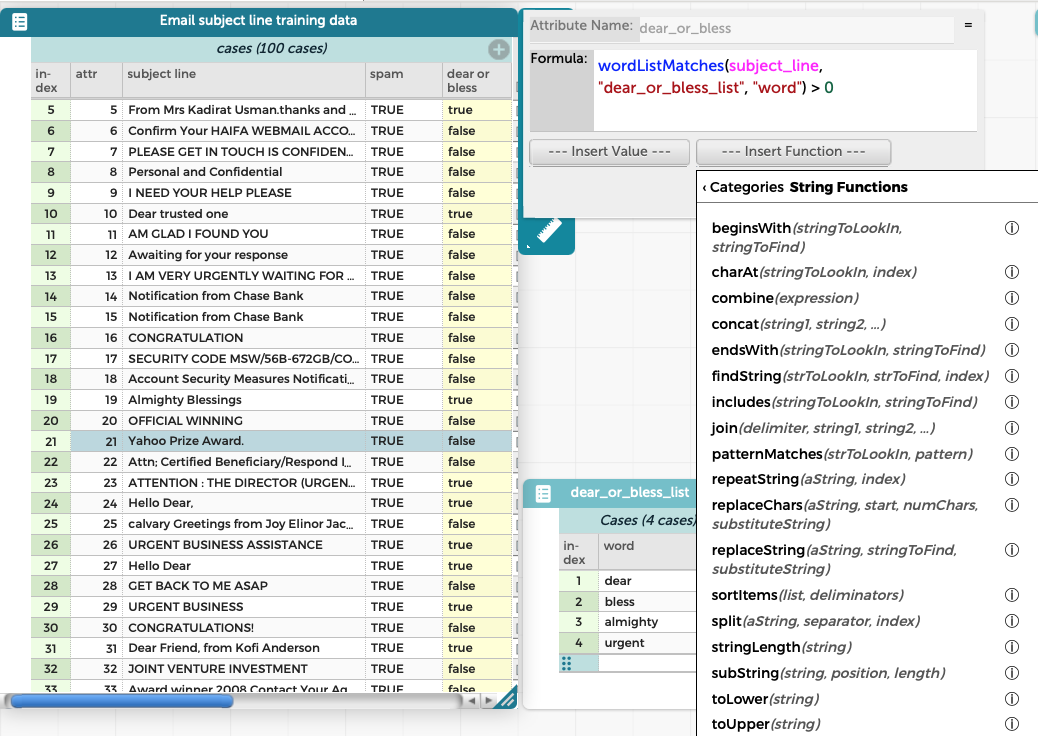} 

}

\caption{Example of creating a new feature (attribute) using a word list. Similar features can be added to the case table using regular expressions or other string functions.}\label{fig:codap08}
\end{figure}

It is straightforward to load the spam MEA subject lines into a CODAP document as a csv file (or access our example documents from the column website).
Once there, it's possible to create new features that might be part of a rule.
Figure \ref{fig:codap08} displays how a new dichotomous attribute (``dear\_or\_bless'') can be added to the case table which is true if the subject line includes any of a set of words (in this case ``dear'', ``bless'', ``almighty'', or ``urgent'').

\begin{figure}

{\centering \includegraphics[width=0.8\linewidth]{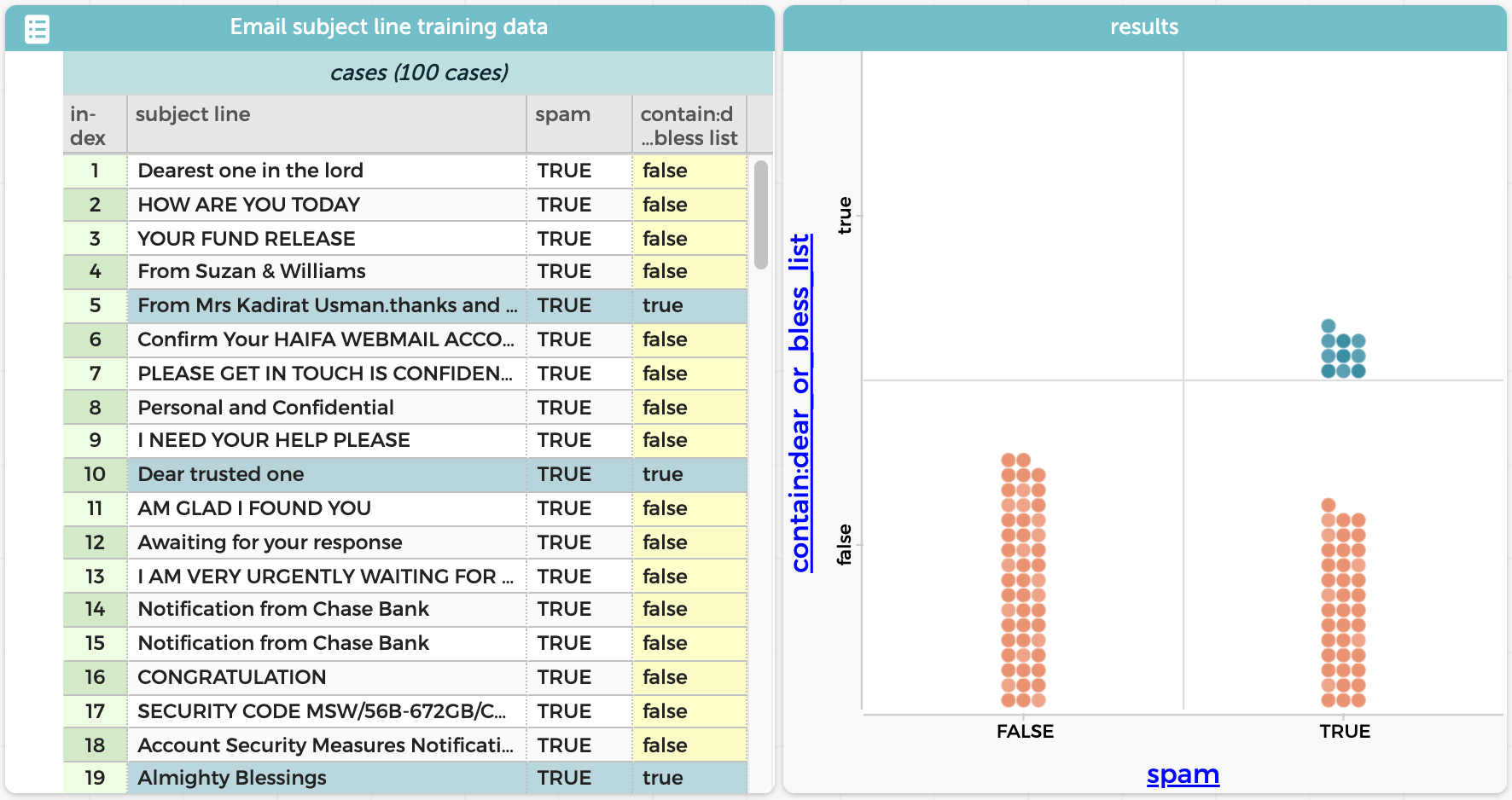} 

}

\caption{All of the subject lines matching words in the "Dear or Bless" list were spam. All non-spams are correctly predicted due to the absence of the words "Dear" or "Bless". In this display, the user has highlighted the spam cell with "Dear or Bless" true, which also highlights those ten subject lines in the case table.}\label{fig:codap09}
\end{figure}

A cross-classification (2x2) table consisting of dots can be created using this feature (on the y-axis) and the true status (spam or not) on the x-axis. Figure \ref{fig:codap09} displays the results when exploring the ``Dear or Bless'' list.

\begin{figure}

{\centering \includegraphics[width=0.8\linewidth]{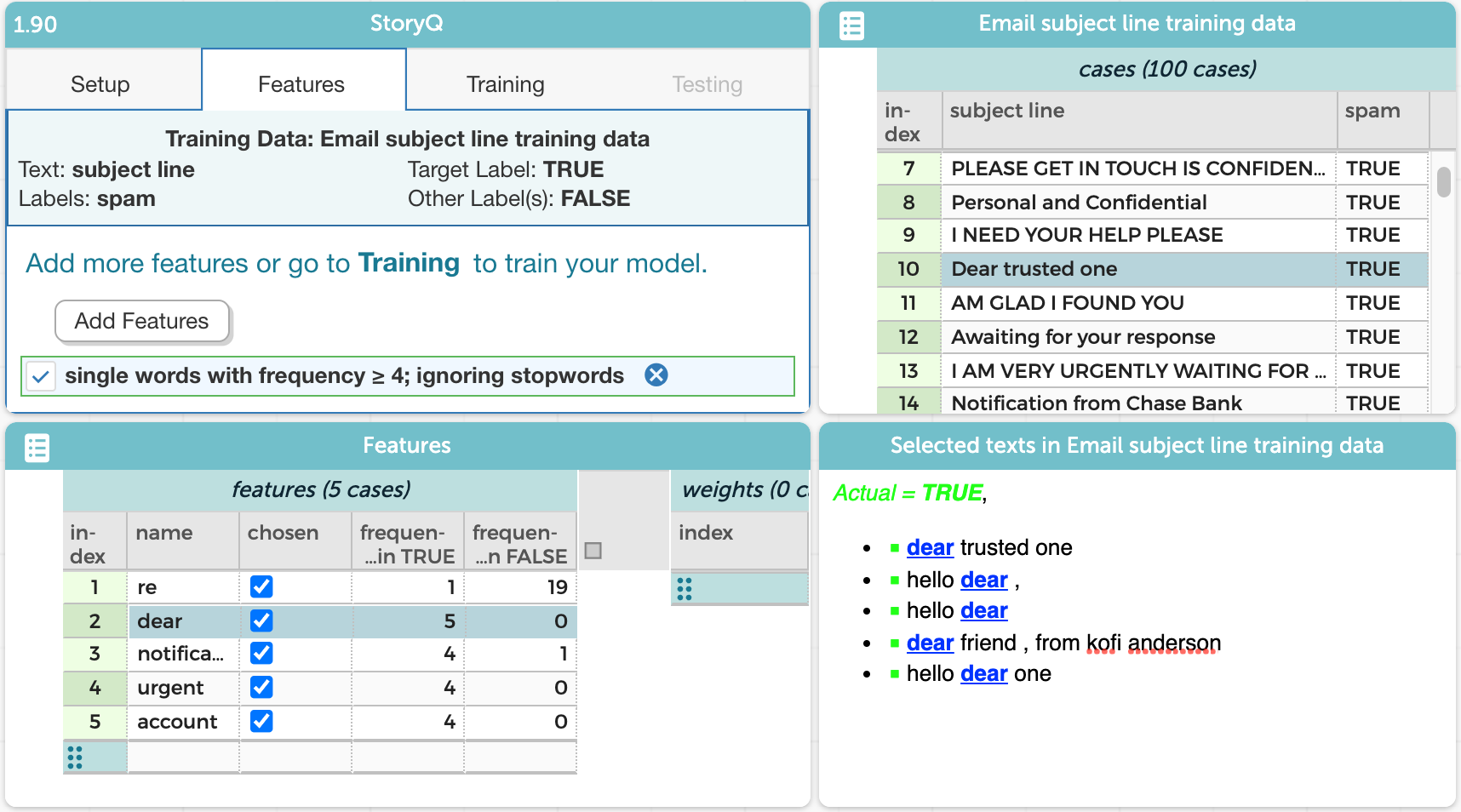} 

}

\caption{StoryQ within CODAP is used to create features using a "bag of words" for features. Certain words seem to be helpful in classifying the subject lines. For example, all five subject lines that include "dear" were spam.}\label{fig:codap03}
\end{figure}

More sophisticated machine learning approaches are straightforward, using extensions to CODAP created for the StoryQ project (Concord Consortium, 2021).
As one example, a ``bag of words'' (or \emph{unigram}) approach can be adopted where every word that appears commonly is added as a dichotomous feature.
Figure \ref{fig:codap03} displays the results once these features are created.
For this example, the words ``Re'', ``Dear'', ``Notification'', ``Urgent'', and ``Account'' occur four or more times and are included as new dichotomous features.

\begin{figure}

{\centering \includegraphics[width=0.8\linewidth]{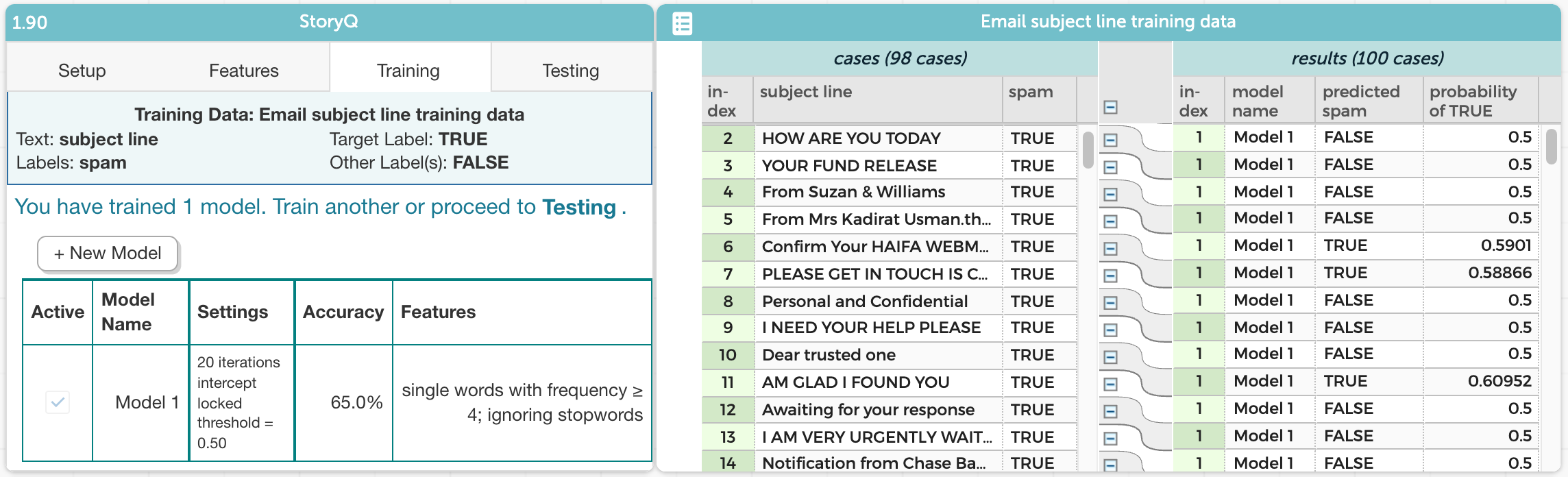} 

}

\caption{StoryQ within CODAP can facilitate training a model to classify spam. In this example, a logistic regression is used with a "bag of words" feature extraction. Predicted probabilities greater than 0.5 were called spam. The model had modest accuracy (0.65), compared to the accuracy of a null model (no predictors) of 0.5 (since exactly half of the data was spam).}\label{fig:codap01}
\end{figure}

A logistic regression can then be fit to predict spam or not using those features.
Figure \ref{fig:codap01} displays the results of the model, where each of the features are used as predictors of whether the subject line is spam.
Any observation with a predicted value greater than 0.5 is classified as spam.

\begin{figure}

{\centering \includegraphics[width=0.8\linewidth]{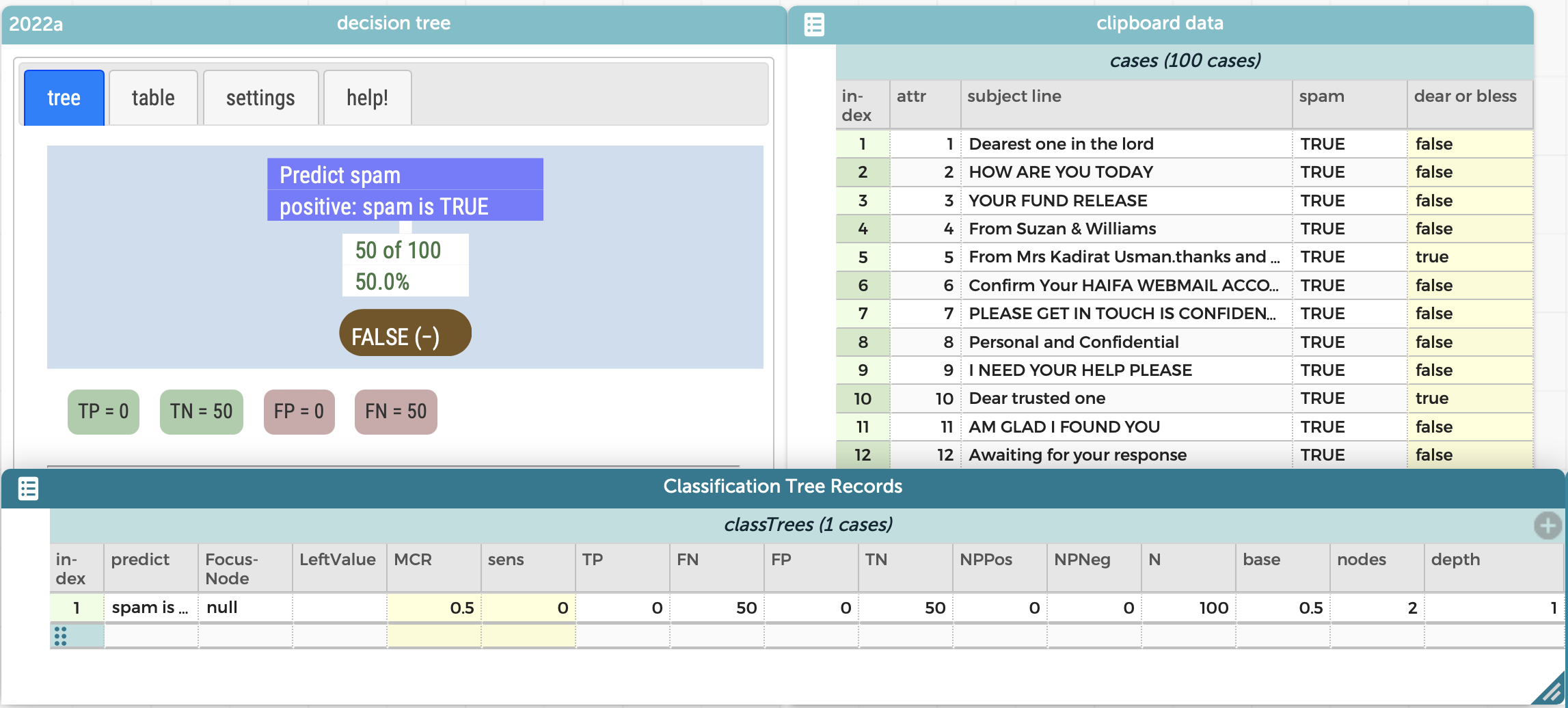} 

}

\caption{The simplest decision tree has no predictors, with all observations declared to be non-spam. The user can specify the outcome of each of the terminal nodes of the tree. A number of summaries are included by default, including misclassification rate (MCR), sensitivity, true positives (TP), false negatives (FN), false positives (FP), and true negatives (TN).}\label{fig:codap10}
\end{figure}

CODAP also features a ``Tree Builder'' which allows the construction of a decision tree to classify spam.
Figure \ref{fig:codap10} displays the simplest decision tree (null model).

A more sophisticated model can be created that splits on whether the words ``dear'' or ``bless'' are included in the subject line as well as whether the string pattern ``Re'' is observed.
Figure \ref{fig:codap01} displays the tree as well as the results from three models (null model, only ``dear or bless'', or the model that incorporates ``dear or bless'' and ``contains\_re'').
Other summaries are available including the traditional \emph{confusion matrix}.

\begin{figure}

{\centering \includegraphics[width=0.8\linewidth]{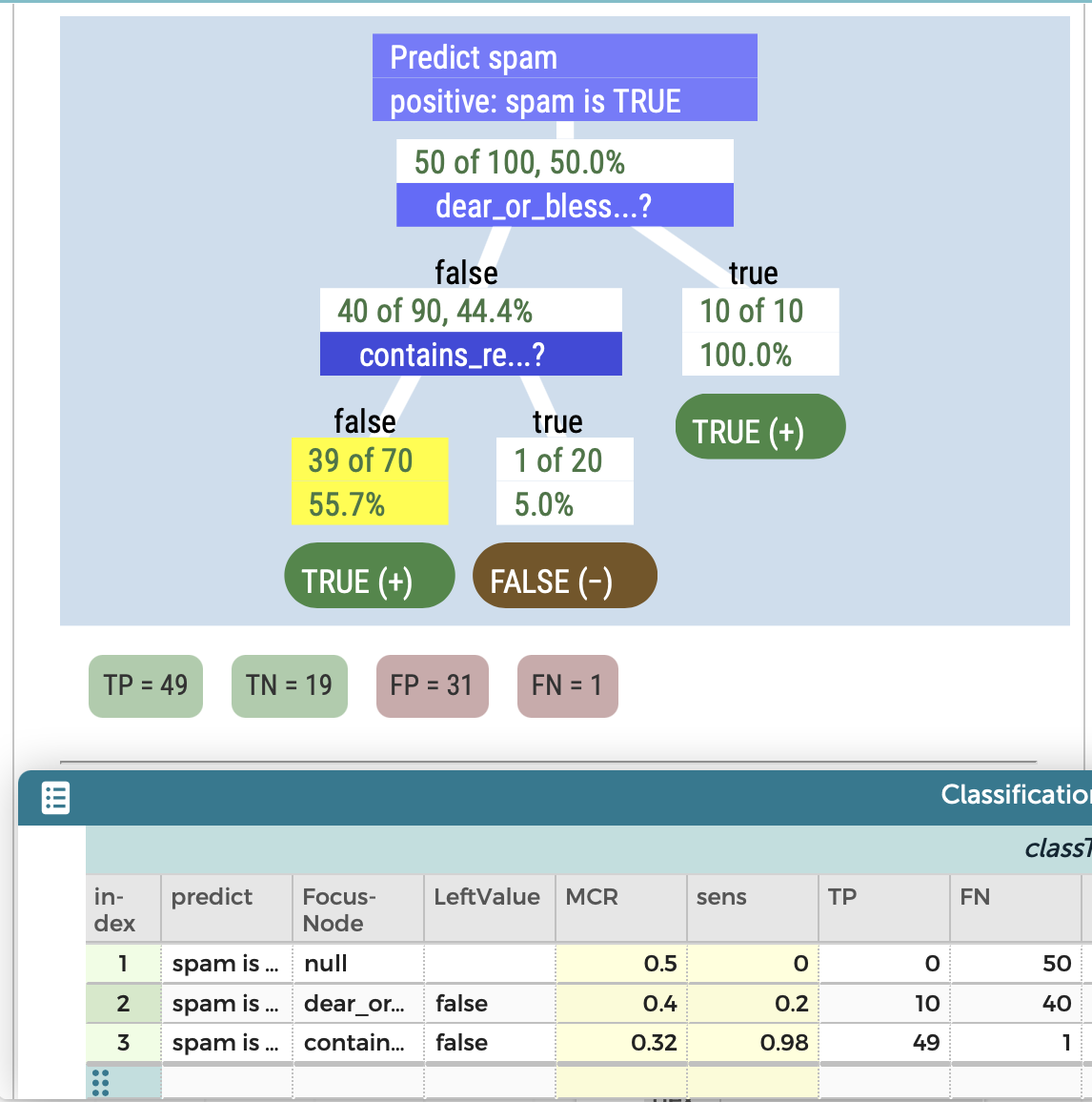} 

}

\caption{A decision which labels 10 subject lines with "dear" or "bless" as spam along with 70 that don't include the string "Re". All twenty remaining messages are labeled as not spam. The model has 68 percent accuracy. Model summaries (including misclassification rate [MCR], 1 - accuracy) are included for three models.}\label{fig:codap13}
\end{figure}

\hypertarget{classifying-spam-with-a-shiny-app}{%
\subsubsection*{Classifying SPAM with a Shiny app}\label{classifying-spam-with-a-shiny-app}}
\addcontentsline{toc}{subsubsection}{Classifying SPAM with a Shiny app}

A third approach to facilitating exploration of classification of subject lines involves the use of a specially designed Shiny web application. The app has been set up with test and training data along with a set of features that could be used to predict whether the subject line is spam or not. Features available in the app include whether the message is all caps, whether it includes a dollar sign, whether it has multiple punctuation marks, whether it includes the string ``Dear'' or ``Mister'', or whether it includes religious subject matter.

\begin{figure}

{\centering \includegraphics[width=0.8\linewidth]{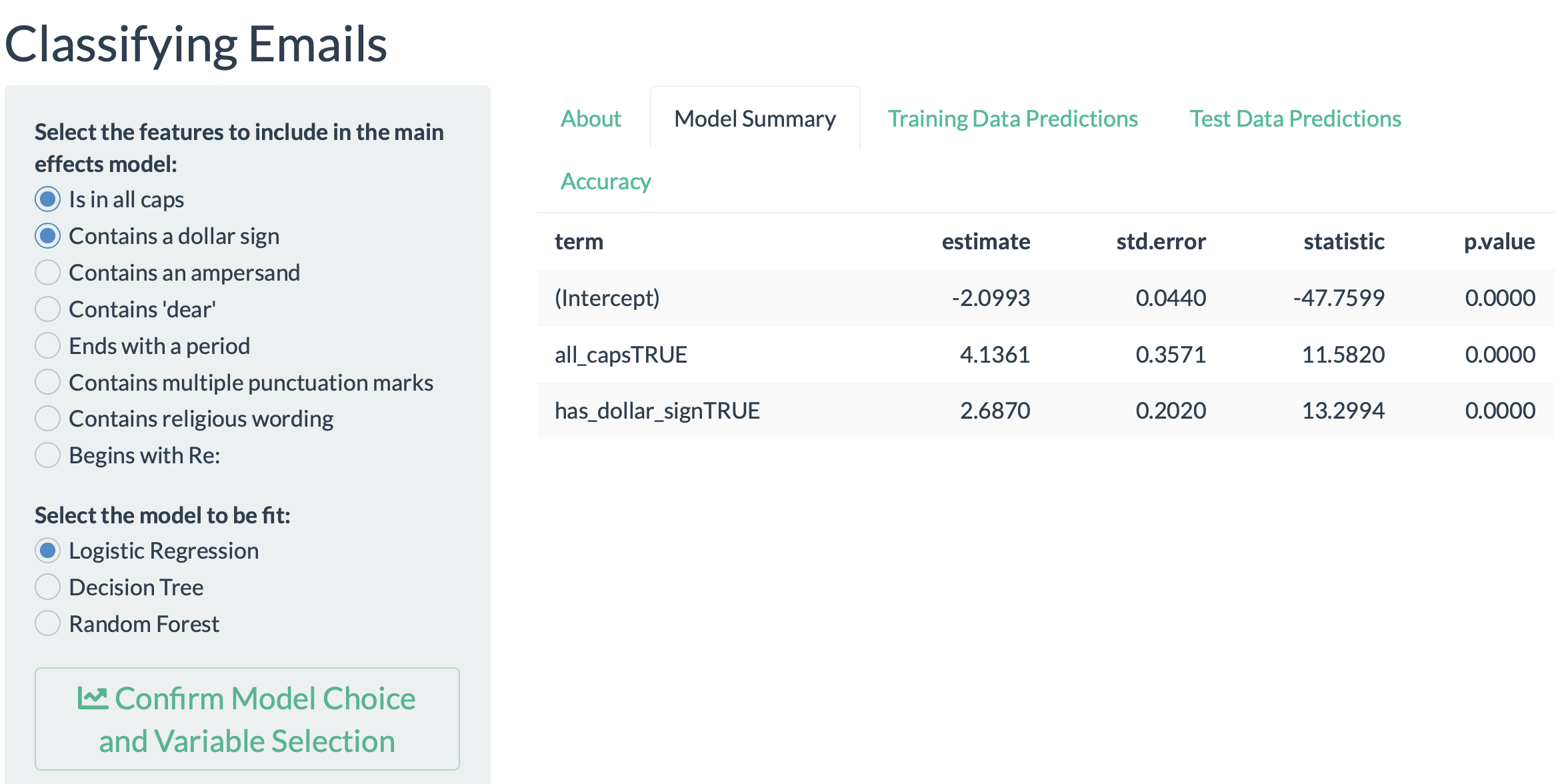} 

}

\caption{A display from the Shiny app where a number of features are used to classify spam subject lines along with the logistic regression coefficients from that specified model. Here two features are included: whether the subject line is all caps and whether the subject line includes a dollar sign.}\label{fig:shiny01}
\end{figure}

\begin{figure}

{\centering \includegraphics[width=0.8\linewidth]{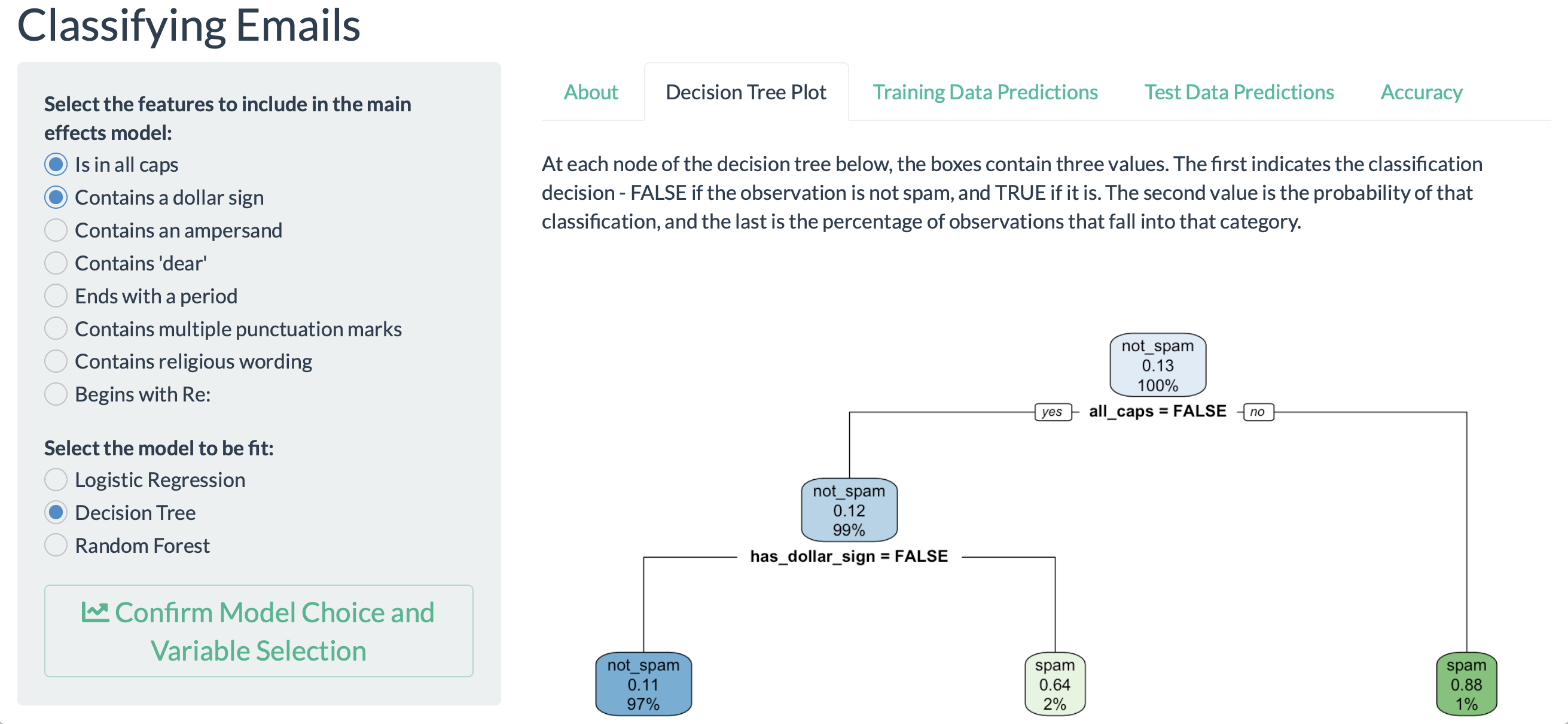} 

}

\caption{A display of the decision tree: the first split depends on whether the entire subject line is all caps: if TRUE the message is marked spam. If not, the next split assesses whether the subject line has a dollar sign. If yes, the message is declared to be spam.}\label{fig:shiny02}
\end{figure}

The user can select which features are to be included and which model (logistic regression, decision tree, or random forest) is to be fit. Tabs allow choice of output including logistic regression coefficients (see Figure \ref{fig:shiny01}), a display of the decision tree (see Figure \ref{fig:shiny02}), predicted values from the model, or a summary of the accuracy of the model on the test and training data.

Users of the app can compare how the accuracy changes when more predictors are added.
They can explore how accuracy differs on the test and training data.
Different directly \emph{interpretable models} (e.g., logistic regression and decision tree) can be compared.
A random forest model can be specified.
This ensemble approach sometimes yields better accuracy, albeit with less interpretability.
These questions provide a useful motivation for a discussion of advantages and disadvantages of classification models.

While constrained to this particular example and feature set, the Shiny app allows the user to fit and interpret three powerful and flexible models without having to install R or write code.
A number of activities might be explored, including calculation of predicted probabilities for a simple logistic regression model, or comparison of the predicted probabilities for a logistic regression vs.~tree model.
No coding is needed for this exploration: as is true for CODAP, all that is required is access to a browser.

Example of the Clickbait Shiny app: \href{https://nicholasjhorton.shinyapps.io/spam_classifier}{\emph{nicholasjhorton.shinyapps.io/spam\_classifier}}

\hypertarget{classifying-spam-using-r}{%
\subsubsection*{Classifying SPAM using R}\label{classifying-spam-using-r}}
\addcontentsline{toc}{subsubsection}{Classifying SPAM using R}

Want more?
There's many ways to extend and further develop approaches to classify SPAM.
One of the case studies in Nolan and Temple Lang's \emph{Data Science in R} (2015) features an extensive exploration of spam classification.
As done in our earlier approaches, they begin by utilizing the subject line and develop and improve a naive Bayes classifier.
Later extensions describe how to ingest other information from the email, such as the header and attachments.
They demonstrate ways to utilize this more extensive feature set, as is often done in commercial spam detection systems.

\hypertarget{conclusion}{%
\subsection*{Conclusion}\label{conclusion}}
\addcontentsline{toc}{subsection}{Conclusion}

In this column we provided an overview of four approaches to engage students in thinking about classifying email messages, a complex problem with non-traditional data.
These varied in their use of technology and student background but all shared a common goal of using data to make better decisions (and how to judge how accurate those decisions were).
We believe that text data deserves a larger space in the curriculum and offer these approaches as tractable ways to get started.

You can find the data, the code for reproducing the figures we presented, and links to the additional resources we mentioned in the GitHub repository for this column at \href{http://bit.ly/taking-a-chance}{\emph{bit.ly/taking-a-chance}}.

\hypertarget{further-reading}{%
\subsection*{Further reading}\label{further-reading}}
\addcontentsline{toc}{subsection}{Further reading}

\begin{enumerate}
\def\labelenumi{\arabic{enumi}.}
\item
  CATALST (Change Agents for Teaching and Learning Statistics) Group (2009). ``Creating a Spam Filter'', \url{https://serc.carleton.edu/sp/cause/mea/examples/example3.html}
\item
  CODAP (Common Online Data Access Platform) (2022). \url{https://codap.concord.org}
\item
  Concord Consortium (2021). ``Narrative Modeling using StoryQ'', \url{https://concord.org/our-work/research-projects/storyq/}
\item
  Nolan, D and Temple Lang, D. (2015). ``Using Statistics to Identify Spam'',
  in ``Data Science in R: A Case Studies Approach to Computational Reasoning and Problem Solving'', CRC Press, \url{https://rdatasciencecases.org}
\item
  Spamlaws.com (2022), ``Spamlaws: How to Stop Scams and Fraud'', \url{https://www.spamlaws.com}
\end{enumerate}

\hypertarget{about-the-authors}{%
\subsection{About the authors}\label{about-the-authors}}

Nicholas J. Horton is Beitzel Professor of Technology and Society (Statistics and Data Science) at Amherst College. He received his doctorate in biostatistics from the Harvard School of Public Health in 1999, and has co-authored a series of books on data science and statistical computing. He is a member of the ASA board of directors, and co-chair of the National Academies Committee on Applied and Theoretical Statistics. This work is part of a larger project with this team that stemmed from Nick's work as a Tinker Fellow with the Concord Consortium.

Jie Chao is a learning scientist at the Concord Consortium.
She received her Ph.D.~in Instructional Technology and STEM education from the University of Virginia in 2012.
Jie is the PI of multiple NSF-funded projects on innovative approaches to STEM teaching and learning.
Her research is focused on designing learning environments that help students develop computational thinking skills, mathematical modeling competencies, and understanding of artificial intelligence.

William Finzer is a Senior Scientist at the Concord Consortium where he leads the development of the Common Online Data Analysis Platform (CODAP).
Bill serves as co-PI on the NSF-funded StoryQ, M2Studio and Boosting Data Fluency projects.
His work centers on bringing data science into the K12 curriculum integrated across subject areas primarily through the creation of data exploration software designed for learning and to be accessible and usable in the classroom.

Phebe Palmer is a recent graduate from Amherst College, having received a B.A. in Statistics in 2021.
Her research centers largely around STEM education, having assisted with projects focused on approaches to statistics pedagogy, as well as equitable access to STEM curriculum.
She works as a research assistant at SageFox Consulting Group based in Amherst, MA.

\end{document}